\documentclass[letterpaper,twocolumn,superscriptaddress,floatfix]{revtex4}
\usepackage[latin1]{inputenc}
\usepackage{bm}
\usepackage{multirow,amssymb,amsbsy,amsmath}
\usepackage{graphicx}
\usepackage{verbatim}
\makeatletter
\usepackage{pifont}
\usepackage{color}
\makeatother

\begin{document}

\title{Coherent dynamics of multi-spin $\rm V_B^-$ centers in hexagonal boron nitride}

\author{Wei Liu}
\thanks{These authors contributed equally to this work.}
\affiliation{CAS Key Laboratory of Quantum Information, University of Science and Technology of China, Hefei, P.R.China}
\affiliation{CAS Center For Excellence in Quantum Information and Quantum Physics, University of Science and Technology of China, Hefei 230026, P.R.China}
\affiliation{Hefei National Laboratory, University of Science and Technology of China, Hefei 230088, China}

\author{Viktor Iv\'{a}dy}
\thanks{These authors contributed equally to this work.}
\affiliation{Max-Planck-Institut f\"{u}r Physik komplexer Systeme, N\"{o}thnitzer Street 38, D-01187 Dresden, Germany}
\affiliation{Department of Physics, Chemistry and Biology, Link\"{o}ping University, SE-581 83 Link\"{o}ping, Sweden}
\affiliation{Wigner Research Centre for Physics, PO Box 49, H-1525, Budapest, Hungary}

\author{Zhi-Peng Li}
\affiliation{CAS Key Laboratory of Quantum Information, University of Science and Technology of China, Hefei, P.R.China}
\affiliation{CAS Center For Excellence in Quantum Information and Quantum Physics, University of Science and Technology of China, Hefei 230026, P.R.China}
\affiliation{Hefei National Laboratory, University of Science and Technology of China, Hefei 230088, China}

\author{Yuan-Ze Yang}
\affiliation{CAS Key Laboratory of Quantum Information, University of Science and Technology of China, Hefei, P.R.China}
\affiliation{CAS Center For Excellence in Quantum Information and Quantum Physics, University of Science and Technology of China, Hefei 230026, P.R.China}
\affiliation{Hefei National Laboratory, University of Science and Technology of China, Hefei 230088, China}

\author{Shang Yu}
\affiliation{CAS Key Laboratory of Quantum Information, University of Science and Technology of China, Hefei, P.R.China}
\affiliation{CAS Center For Excellence in Quantum Information and Quantum Physics, University of Science and Technology of China, Hefei 230026, P.R.China}
\affiliation{Hefei National Laboratory, University of Science and Technology of China, Hefei 230088, China}

\author{Yu Meng}
\affiliation{CAS Key Laboratory of Quantum Information, University of Science and Technology of China, Hefei, P.R.China}
\affiliation{CAS Center For Excellence in Quantum Information and Quantum Physics, University of Science and Technology of China, Hefei 230026, P.R.China}
\affiliation{Hefei National Laboratory, University of Science and Technology of China, Hefei 230088, China}

\author{Zhao-An Wang}
\affiliation{CAS Key Laboratory of Quantum Information, University of Science and Technology of China, Hefei, P.R.China}
\affiliation{CAS Center For Excellence in Quantum Information and Quantum Physics, University of Science and Technology of China, Hefei 230026, P.R.China}
\affiliation{Hefei National Laboratory, University of Science and Technology of China, Hefei 230088, China}

\author{Nai-Jie Guo}
\affiliation{CAS Key Laboratory of Quantum Information, University of Science and Technology of China, Hefei, P.R.China}
\affiliation{CAS Center For Excellence in Quantum Information and Quantum Physics, University of Science and Technology of China, Hefei 230026, P.R.China}
\affiliation{Hefei National Laboratory, University of Science and Technology of China, Hefei 230088, China}

\author{Fei-Fei Yan}
\affiliation{CAS Key Laboratory of Quantum Information, University of Science and Technology of China, Hefei, P.R.China}
\affiliation{CAS Center For Excellence in Quantum Information and Quantum Physics, University of Science and Technology of China, Hefei 230026, P.R.China}
\affiliation{Hefei National Laboratory, University of Science and Technology of China, Hefei 230088, China}

\author{Qiang Li}
\affiliation{CAS Key Laboratory of Quantum Information, University of Science and Technology of China, Hefei, P.R.China}
\affiliation{CAS Center For Excellence in Quantum Information and Quantum Physics, University of Science and Technology of China, Hefei 230026, P.R.China}
\affiliation{Hefei National Laboratory, University of Science and Technology of China, Hefei 230088, China}

\author{Jun-Feng Wang}
\affiliation{CAS Key Laboratory of Quantum Information, University of Science and Technology of China, Hefei, P.R.China}
\affiliation{CAS Center For Excellence in Quantum Information and Quantum Physics, University of Science and Technology of China, Hefei 230026, P.R.China}
\affiliation{Hefei National Laboratory, University of Science and Technology of China, Hefei 230088, China}

\author{Jin-Shi Xu}
\affiliation{CAS Key Laboratory of Quantum Information, University of Science and Technology of China, Hefei, P.R.China}
\affiliation{CAS Center For Excellence in Quantum Information and Quantum Physics, University of Science and Technology of China, Hefei 230026, P.R.China}
\affiliation{Hefei National Laboratory, University of Science and Technology of China, Hefei 230088, China}

\author{Xiao Liu}
\affiliation{CAS Key Laboratory of Quantum Information, University of Science and Technology of China, Hefei, P.R.China}
\affiliation{CAS Center For Excellence in Quantum Information and Quantum Physics, University of Science and Technology of China, Hefei 230026, P.R.China}
\affiliation{Hefei National Laboratory, University of Science and Technology of China, Hefei 230088, China}

\author{Zong-Quan Zhou}
\affiliation{CAS Key Laboratory of Quantum Information, University of Science and Technology of China, Hefei, P.R.China}
\affiliation{CAS Center For Excellence in Quantum Information and Quantum Physics, University of Science and Technology of China, Hefei 230026, P.R.China}
\affiliation{Hefei National Laboratory, University of Science and Technology of China, Hefei 230088, China}

\author{Yang Dong}
\affiliation{CAS Key Laboratory of Quantum Information, University of Science and Technology of China, Hefei, P.R.China}
\affiliation{CAS Center For Excellence in Quantum Information and Quantum Physics, University of Science and Technology of China, Hefei 230026, P.R.China}
\affiliation{Hefei National Laboratory, University of Science and Technology of China, Hefei 230088, China}

\author{Xiang-Dong Chen}
\affiliation{CAS Key Laboratory of Quantum Information, University of Science and Technology of China, Hefei, P.R.China}
\affiliation{CAS Center For Excellence in Quantum Information and Quantum Physics, University of Science and Technology of China, Hefei 230026, P.R.China}
\affiliation{Hefei National Laboratory, University of Science and Technology of China, Hefei 230088, China}

\author{Fang-Wen Sun}
\affiliation{CAS Key Laboratory of Quantum Information, University of Science and Technology of China, Hefei, P.R.China}
\affiliation{CAS Center For Excellence in Quantum Information and Quantum Physics, University of Science and Technology of China, Hefei 230026, P.R.China}
\affiliation{Hefei National Laboratory, University of Science and Technology of China, Hefei 230088, China}

\author{Yi-Tao Wang}
\email{yitao@ustc.edu.cn}
\affiliation{CAS Key Laboratory of Quantum Information, University of Science and Technology of China, Hefei, P.R.China}
\affiliation{CAS Center For Excellence in Quantum Information and Quantum Physics, University of Science and Technology of China, Hefei 230026, P.R.China}
\affiliation{Hefei National Laboratory, University of Science and Technology of China, Hefei 230088, China}

\author{Jian-Shun Tang}
\email{tjs@ustc.edu.cn}
\affiliation{CAS Key Laboratory of Quantum Information, University of Science and Technology of China, Hefei, P.R.China}
\affiliation{CAS Center For Excellence in Quantum Information and Quantum Physics, University of Science and Technology of China, Hefei 230026, P.R.China}
\affiliation{Hefei National Laboratory, University of Science and Technology of China, Hefei 230088, China}

\author{Adam Gali}
\email{gali.adam@wigner.hu}
\affiliation{Wigner Research Centre for Physics, PO Box 49, H-1525, Budapest, Hungary}
\affiliation{Department of Atomic Physics, Institute of Physics, Budapest University of Technology and Economics, M\H{u}egyetem rakpart 3., H-1111 Budapest, Hungary}

\author{Chuan-Feng Li}
\email{cfli@ustc.edu.cn}
\affiliation{CAS Key Laboratory of Quantum Information, University of Science and Technology of China, Hefei, P.R.China}
\affiliation{CAS Center For Excellence in Quantum Information and Quantum Physics, University of Science and Technology of China, Hefei 230026, P.R.China}
\affiliation{Hefei National Laboratory, University of Science and Technology of China, Hefei 230088, China}

\author{Guang-Can Guo}
\affiliation{CAS Key Laboratory of Quantum Information, University of Science and Technology of China, Hefei, P.R.China}
\affiliation{CAS Center For Excellence in Quantum Information and Quantum Physics, University of Science and Technology of China, Hefei 230026, P.R.China}
\affiliation{Hefei National Laboratory, University of Science and Technology of China, Hefei 230088, China}

\begin{abstract}

Hexagonal boron nitride (hBN) has recently been demonstrated to contain optically polarized and detected electron spins that can be utilized for implementing qubits and quantum sensors in nanolayered-devices. Understanding the coherent dynamics of microwave driven spins in hBN is of crucial importance for advancing these emerging new technologies. Here, we demonstrate and study the Rabi oscillation and related dynamical phenomena of the negatively charged boron vacancy ($\rm V_B^-$) spins in hBN. We report on different dynamics of the $\rm V_B^-$ spins at weak and strong magnetic fields. In the former case the defect behaves like a single electron spin system, while in the latter case it behaves like a multi-spin system exhibiting the multiple-frequency dynamical oscillation like clear beat in Ramsey fringes. We also carry out theoretical simulations for the spin dynamics of $\rm V_B^-$ and reveal that the nuclear spins can be driven via the strong electric-nuclear coupling existing in $\rm V_B^-$ center, which can be modulated by the magnetic field and microwave field.

\end{abstract}

\maketitle
\date{\today}

Van der Waals (vdW) materials exhibit diverse electronic properties from semimetal (e.g., graphene) through semiconductor (e.g., transition metal dichalcogenides, TMDCs for short) to insulator (e.g., hexagonal boron nitride, or hBN) \cite{xia2014two}. Their common feature is the layered structure, namely, the intralayer atoms are combined by strong chemical bonds while the interlayer is connected by the relatively weak vdW force. This feature makes the layers of different vdW materials easy to be stacked together to form heterostructures \cite{geim2013van}, which have the advantage of no lattice mismatch compared to their three-dimensional counterparts, including GaAs, silicon, or diamond, etc. Besides, vdW materials have strong interaction with light since the two-dimensional confinement of the electronic states \cite{trovatello2021optical}. These characteristics lead to a great variety of applications of vdW materials, such as photocurrent generation \cite{yu2013highly}, light-emitting diode \cite{ross2014electrically}, field effect transistor \cite{liu2013electrically}, single photon \cite{srivastava2015optically, he2015single, koperski2015single, chakraborty2015voltage, tonndorf2015single, palacios2017large, branny2017deterministic, errando2020on, tran2016Quantum, tran2016quantum, tran2016robust, martinez2016efficient, chejanovsky2016structural, choi2016engineering, grosso2017tunable, xue2018anomalous, proscia2018near, liu2020an, fournier2020position, barthelmi2020atomistic} and optical parametric amplification \cite{trovatello2021optical}. Moreover, the light-valley interaction in TMDCs leads to the field of valleytronics \cite{xu2014spin, manzeli20172d}. All these applications will expectedly contribute to the design and construction of photonic and electronic devices in nanoscale, benefited from the atomic thickness of vdW materials.

Among this family of layered materials, hBN has a large bandgap of $\sim$6 eV, which makes it possible to host plenty of defect states in bandgap similar to diamond \cite{barry2020sensitivity,hanson2008coherent,chen2015subdiffraction} and silicon carbide \cite{wang2020coherent,yan2020room,li2021room}. Single-layer hBN was first found to host room-temperature quantum emitter in 2016 by Tran \emph{et al.} \cite{tran2016Quantum}, that stimulated numerous works to explore promising quantum emitters \cite{tran2016quantum, tran2016robust, martinez2016efficient, chejanovsky2016structural, choi2016engineering, grosso2017tunable, xue2018anomalous, proscia2018near, liu2020an, fournier2020position} and potential solid spin qubits \cite{exarhos2019magnetic, toledo2018electron, gottscholl2020initialization, chejanovsky2021single, mendelson2021identifying,kianinia2020generation,gao2021femtosecond,liu2021temperature,gottscholl2021room}. Quantum emitters in hBN (in monolayer, flake or bulk) have the advantages of high brightness \cite{grosso2017tunable, liu2020an}, broad spectral range \cite{tran2016robust}, easy tunability \cite{grosso2017tunable, xue2018anomalous}, and easy fabrication. A diverse set of fabrication methods has been demonstrated like chemical etching \cite{chejanovsky2016structural}, electron or ion irradiation \cite{chejanovsky2016structural, choi2016engineering,kianinia2020generation,fournier2020position}, laser ablation \cite{choi2016engineering,gao2021femtosecond}, strain \cite{proscia2018near}.

\begin{figure*}[tb]
\centering
\includegraphics[width=0.8\textwidth]{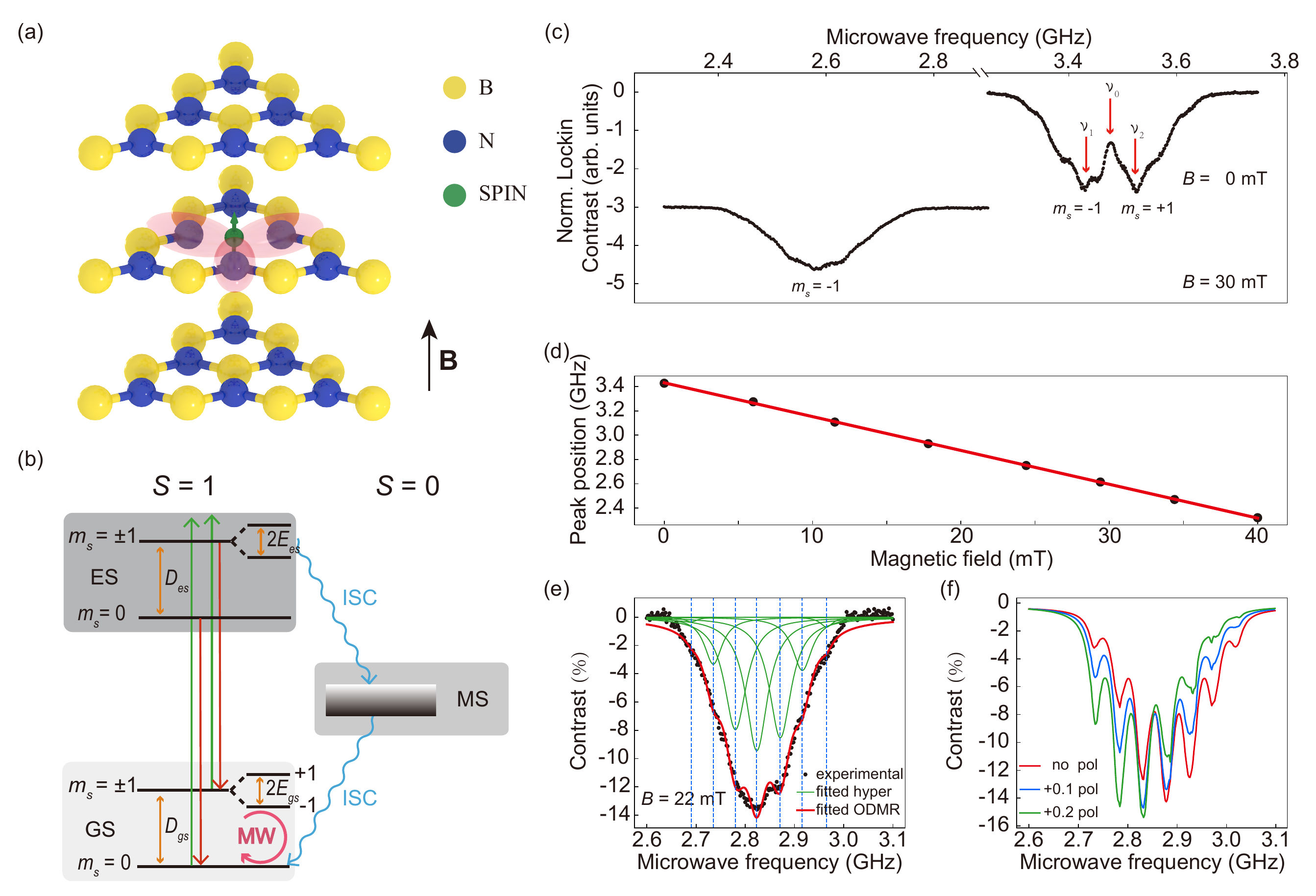}
\caption{Simplified atomic structure, energy levels and ODMR results of $\rm V_B^-$ center. (a) Geometrical structure of $\rm V_B^-$ defect in hBN with alternating boron (yellow), nitrogen (blue) atoms and a negatively charged boron vacancy (green arrows). The negatively
charged boron vacancy comprises a missing boron atom and an extra electron surrounded by three equivalent nitrogen atoms. The magnetic field applied in the experiment is perpendicular to the two-dimensional plane of hBN, that is, parallel to the hexagonal $c$ axis. (b) Simplified energy levels of $\rm V_B^-$ center and the related optical transitions among ground states (GS), excited states (ES) and metastable states (MS). The 532-nm laser (green) is used for the spin polarization and readout, and the microwave (pink) is used for coherent control of the spin state. (c) Room-temperature ODMR spectra measured at 0-mT (top pane) and 30-mT (lower) magnetic fields. At 0 mT, the ODMR spectrum 
is fitted by a two Lorentzian function to obtain the energy-level splittings $\nu_{1} \sim$3.424 GHz and $\nu_{2} \sim$3.533 GHz.
%exhibit obvious hyperfine structures, which indicate the nucleus-electron interaction between $\rm V_B^-$ and the three neighboring nitrogen nuclei ($^{14}$N). 
(d) Dependence of the $m_s=-1\leftrightarrow m_s=0$ splitting shift on magnetic field, from which we obtain the $g$ factor of $\rm V_B^-$ spin to be $1.992\pm0.010$. (e) Hyperfine structure of the ODMR spectrum measured at $B = 22$ mT, fitted with a seven Lorentzian function (solid lines). The dashed lines marked the seven hyperfine peaks with a characteristic splitting of $A \sim 45.8$ MHz. (f) Theoretical ODMR spectra as $B=21.7$ mT for nonpolarized (red) and polarized (blue and green) nearest neighbor $^{14}$N nuclear spins, where label ``$x$ pol'' means the nuclear populations are 1/3+$x$, 1/3, and 1/3-$x$ on the $m_\text{I}$ = +1, 0, and -1 nuclear-spin states, respectively.}
\label{Fig1}
\end{figure*}

Recently, defects in hBN have attracted a lot of attentions as a good candidate for solid spin qubit (particularly, in the vdW-nano-devices). Electron paramagnetic resonance (EPR) signals in hBN have been found in very early decades \cite{katzir1975point, moore1972electron, fanciulli1992wide}, whose origin has been identified by numerical calculations recently \cite{sajid2018defect, abdi2018color}. The theoretical works have also predicted many possible defects in hBN to give rise to optically detected magnetic resonance (ODMR) signals. In experiment, Exarhos \emph{et al.} \cite{exarhos2019magnetic} found the magnetic-field-dependent fluorescence intensity of a hBN defect in 2019. Later, related ODMR signals were revealed by Gottscholl \emph{et al.} \cite{toledo2018electron,gottscholl2020initialization}, Chejanovsky \emph{et al.} \cite{chejanovsky2021single} and Mendelson \emph{et al.} \cite{mendelson2021identifying}, the defects were tentatively assigned to be $\rm V_B^-$ (negatively charged boron vacancy) defects \cite{toledo2018electron,gottscholl2020initialization} or the defects related to carbon \cite{mendelson2021identifying}. After these initial experimental results, several theoretical analyzations were carried out, especially for $\rm V_B^-$ defects \cite{ivady2020ab, sajid2020edge}, and the temperature-dependent features of $\rm V_B^-$ spin defect have also been investigated experimentally \cite{liu2021temperature}. Here, we demonstrate the room-temperature coherent control of $\rm V_B^-$ spin in hBN. Especially, the Rabi oscillations under different magnetic fields indicate the strong electronic-nuclear spin coupling existing in $\rm V_B^-$ center. We also 
carry out the theoretical simulations of $\rm V_B^-$ spin dynamics, and reveal that the nuclear spins can be modulated by the microwave (MW) and magnetic field, further give rise to the multi-spin dynamics of $\rm V_B^-$ center.

\begin{figure*}[tb]
\centering
\includegraphics[width=0.95\textwidth]{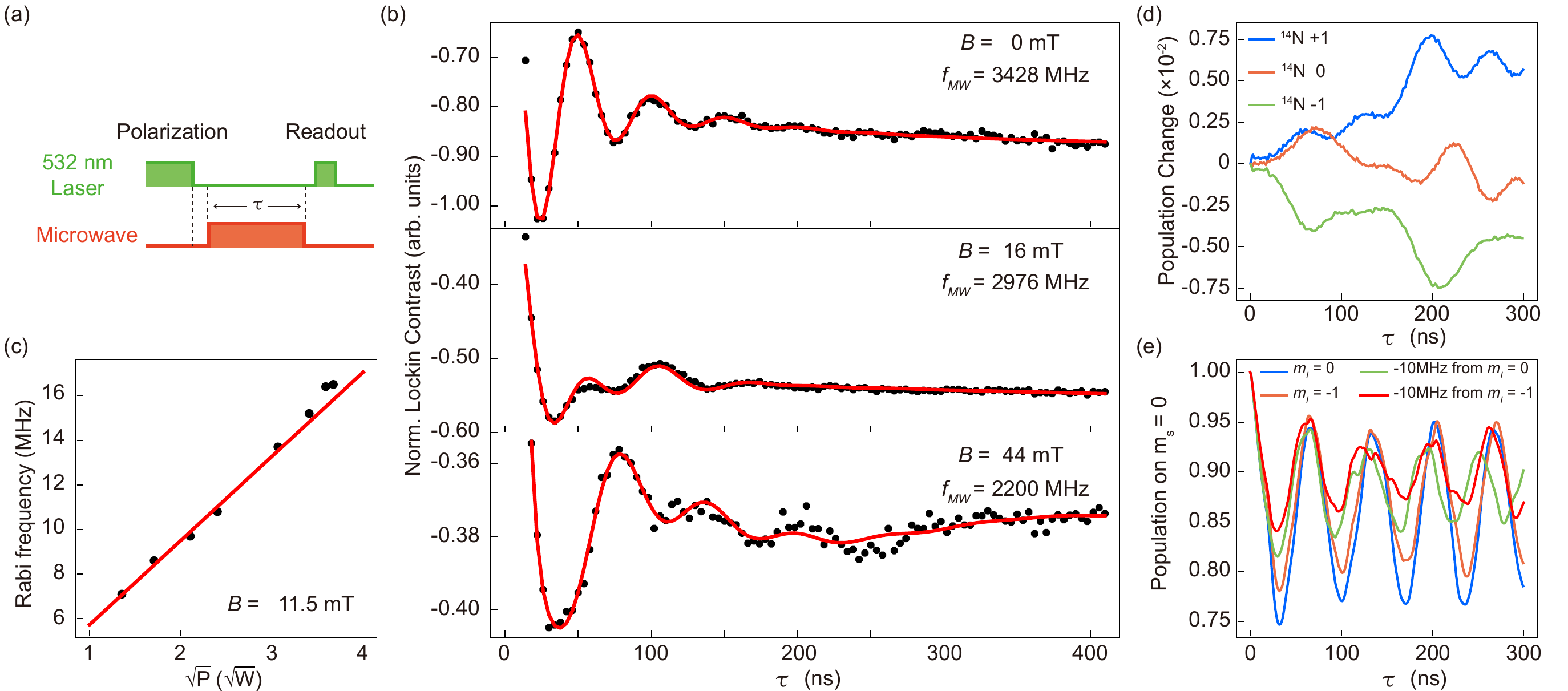}
\caption{Rabi oscillations. (a) Pulse sequence of Rabi measurement comprising of a first laser pulse for spin polarization, then a microwave pulse with length $\tau$ for spin manipulation and a second laser pulse for state readout. (b) Rabi oscillations on the $m_s=-1\leftrightarrow m_s=0$ transition observed at room temperature and different magnetic fields. The data are fitted using $a e^{-\tau/T_a}\prod_{i=1}^{n}\cos(2\pi f_i\tau+\phi_i)+b e^{-\tau/T_b}+c$, with the amount of the different oscillation components $n = 1$ for $B = 0$ mT, $n=2$ for $B = 16$, and $n=3$ for $B = 44$ mT (red lines). The fitting parameters $a$, $T_a$, $f_i$, $\phi_i$, $b$, $T_b$ and $c$ are amplitude, oscillation decay time, frequency, phase, amplitude and background decay time and constant background, respectively. (c) Linear dependence of Rabi frequency on the square root of microwave power $\sqrt{P}$. (d) Theoretical dynamical oscillation of $^{14}$N nuclear spins with the driving MW resonant at the $m_I=-1$ hyperfine peak. The blue, red and green curves show the dynamical polarization change on the $m_\text{I}$ = +1, 0, and -1 nuclear spin states under continuous MW driving, respectively.
(e) Theoretical Rabi oscillation of the 4-spin $\rm V_B^-$ system. The blue, orange, green and red curves show Rabi oscillations driven by the MW with frequencies at center hyperfine peak, $m_I=-1$ hyperfine peak, -10 MHZ detuning from $m_I=0$ hyperfine peak, and -10 MHz detuning from $m_I=-1$ hyperfine peak, respectively.}
\label{Fig2}
\end{figure*}

\section*{Results}
\textbf{Sample Description.} The hBN sample we investigated here is a bulk synthetic monocrystalline flake with the $\sim$1 mm lateral size purchased from HQ Graphene. The sample is irradiated by neutron flow with a thermal flux of 1.4$\times$10$^{13}$ cm$^{-2}$s$^{-1}$ for 4 hours in a nuclear reactor (integrated dose $\sim$2$\times$10$^{17}$ cm$^{-2}$), similar to that in Ref. \cite{toledo2018electron} to generate the $\rm V_B^-$ defects. 
%The irradiated hBN crystals apper pink \cite{toledo2018electron}, 
The photoluminescence (PL) of the irradiated hBN flake is characterized by a broadband spectrum centering around $\sim$800 nm (see Supplementary
Information S2), that is consistent with the previously reported theoretical and experimental PL spectra of $\rm V_B^-$ defects \cite{reimers2020photoluminescence,gottscholl2020initialization,ivady2020ab}. %measured in our home-built confocal microscopy system (see Methods).

The atomic structure of $\rm V_B^-$ defect is shown in Fig. 1(a), and the simplified diagram of the energy levels of $\rm V_B^-$ spin is sketched in Fig. 1(b). As discussed in Refs. \cite{gottscholl2020initialization,ivady2020ab,reimers2020photoluminescence}, $\rm V_B^-$ defect contains a triplet ground state (GS), which is primarily constituted of three energy levels with $m_s=0$ and $m_s=\pm1$, and $D$ is the zero-field splitting (ZFS) between them. ES and MS represent the triplet ($S = 1$) excited state and the metastable singlet (S=0) state, respectively. The green arrows represent the laser excitation which pump the population to ES, and the red arrows represent the fluorescence to be detected. The blue wavy arrows represent inter-system crossings (ISC) between $S=1$ and $S=0$ states. The pink circled arrow is the applied microwave (MW) between $m_s=0$ and $m_s=\pm1$, which will modulate the populations of these states, and hence change the intensity of the fluorescence. By recording the difference of the fluorescence intensities, we can read out the state of the spin qubit. This method is called ODMR. We perform the ODMR measurement to further confirm the origin and character of the generated spin defect in our hBN sample.

\textbf{ODMR results.} By sweeping the frequency of the MW field, typical ODMR spectra of the $\rm V_B^-$ in hBN at zero and 30-mT magnetic fields are obtained, see Fig. 1(c). Here the excitation laser is always on and the MW operates at on/off mode, then the contrast is obtained from the difference between the MW-on and MW-off fluorescences. The frequency $\nu_0=(\nu_1+\nu_2)/2=3.479$ GHz corresponds to the ZFS $D$, where $\nu_1$ ($\nu_2$) corresponds to the energy-level splittings between $m_s=-1$ ($m_s=+1$) and $m_s=0$ states. We also observe the frequency shift with the magnetic field for the $m_s=-1\leftrightarrow m_s=0$ transition and fit the $g$ factor of this spin to be $1.992 \pm 0.010$ as shown in Fig. 1(d) (See more ODMR spectra in Supplementary Information S3). Remarkably, for each transition peak of $m_s=\pm1$ to $m_s=0$, we can clearly see several hyperfine peaks, which should be related to the electronic-nuclear hyperfine interaction.

\begin{figure*}[tb]
\centering
\includegraphics[width=0.75\textwidth]{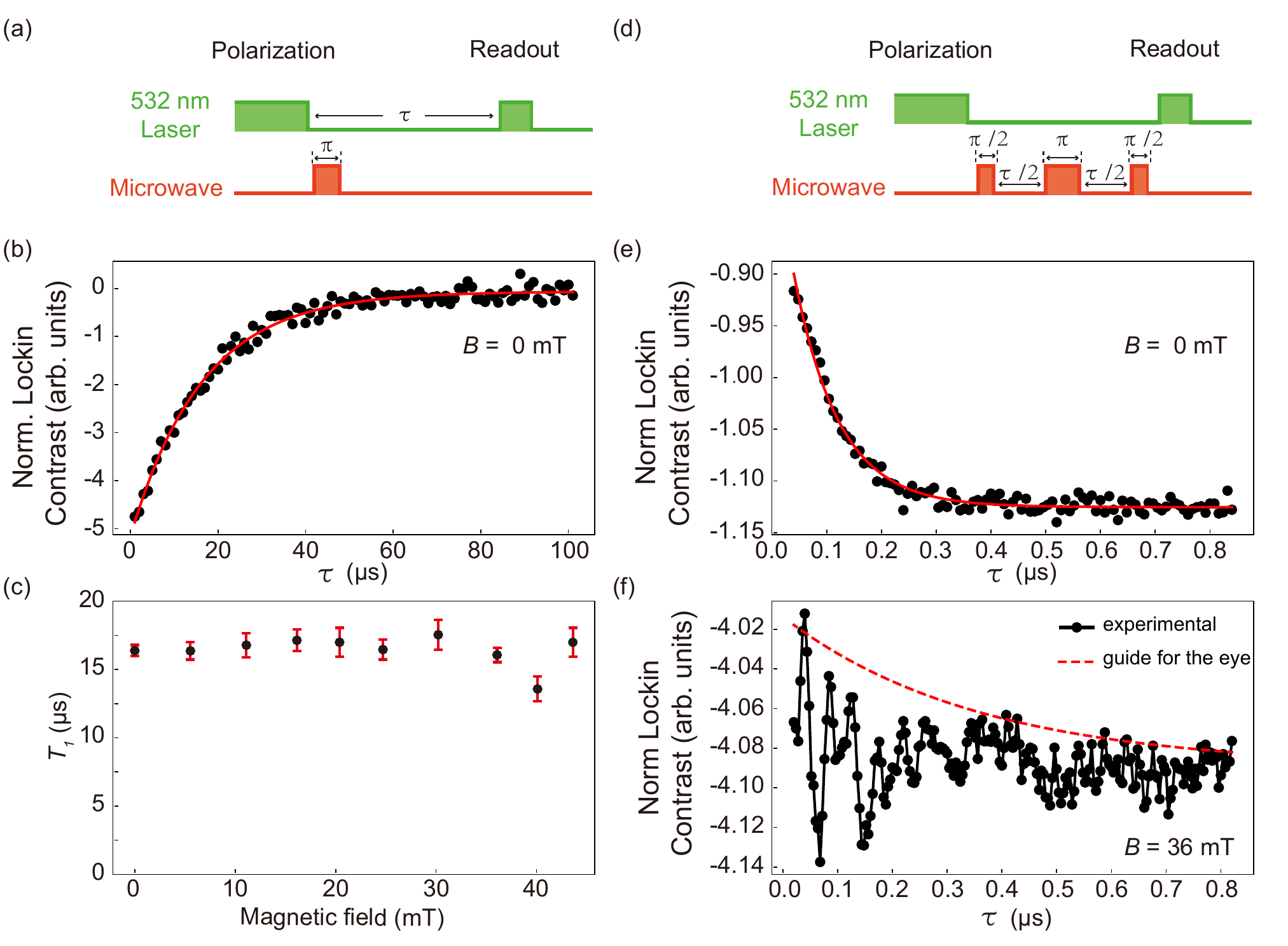}
\caption{$T_1$ measurement and spin echo detections. (a) Pulse sequence for characterizing the spin-lattice relaxation dynamics, including the spin polarization and readout laser pulse, the $\pi$-pulse microwave obtained from Rabi measurement and the varying free evolution time $\tau$. (b) $T_1$ measurement at 0 mT revealing the spin-lattice relaxation time of $T_1 = 16.377 \pm 0.416$ $\mu$s. (c) $T_1$ time versus magnetic field, suggesting that there is roughly no $T_1$ dependence on magnetic field. (d) Pulse sequence for spin echo measurement with $\frac{\pi}{2}-\frac{\tau}{2}-\pi-\frac{\tau}{2}-\frac{\pi}{2}$ sequence, where $\tau$ is the free evolution time. (e) Optically-detected spin-echo measurement at 0 mT. (f) Spin echo at 36 mT which cannot be fitted well, showing complicated oscillations induced by the nuclear spin bath and the red line is only a guide for eyes.}
\label{Fig3}
\end{figure*}

\textbf{Hyperfine structures.} For the $\rm V_B^-$ center in hBN, there are three nearest neighbor nitrogen nuclei ($^{14}$N) surrounding the boron vacancy site, each of which has a nuclear spin of $I=1$. Then the electron spin and the three $^{14}$N nuclear spins are coupled and form a 4-spin system whose atomic structure is marked by the red lobes in Fig. 1(a), and a total of $2(3I)+1=7$ hyperfine transitions should be observed in the ODMR spectra. To identify this, we carry out the theoretical simulation for the 4-spin $\rm V_B^-$ system, and obtain the theoretical ODMR spectra at 21.7-mT magnetic field. The theoretical ODMR spectra exhibit seven distinct hyperfine peaks with the hyperfine splitting of $\sim48$ MHz. The polarization of $^{14}$N nuclear spin can generate the asymmetric ODMR spectra as shown in Fig. 1(f). Further, we fit the experimental ODMR spectra with seven Lorentzian functions as shown in Fig. 1(e). The experimental hyperfine splitting is $\sim45$ MHz, which is consistent with our theoretical results and the previously reported works of $\rm V_B^-$ spin \cite{gottscholl2020initialization,ivady2020ab}. Considering the fact that the ODMR hyperfine structure depends on the intrinsic electronic-nuclear structure, the generated defects in this work can be identified as the $\rm V_B^-$ defect. In addition, the observed ODMR spectra in Fig. (c) and (e) also exhibit asymmetric hyperfine structures indicating the polarization of $^{14}$N nuclear spins occurring in $\rm V_B^-$ center under the MW drive. By studying fitted hyperfine structures of more ODMR spectra at different magnetic fields, we find that the $^{14}$N-spin polarization is enhanced with the increase of external magnetic field (see Supplementary Information S3). 

\begin{figure*}[tb]
\centering
\includegraphics[width=0.95\textwidth]{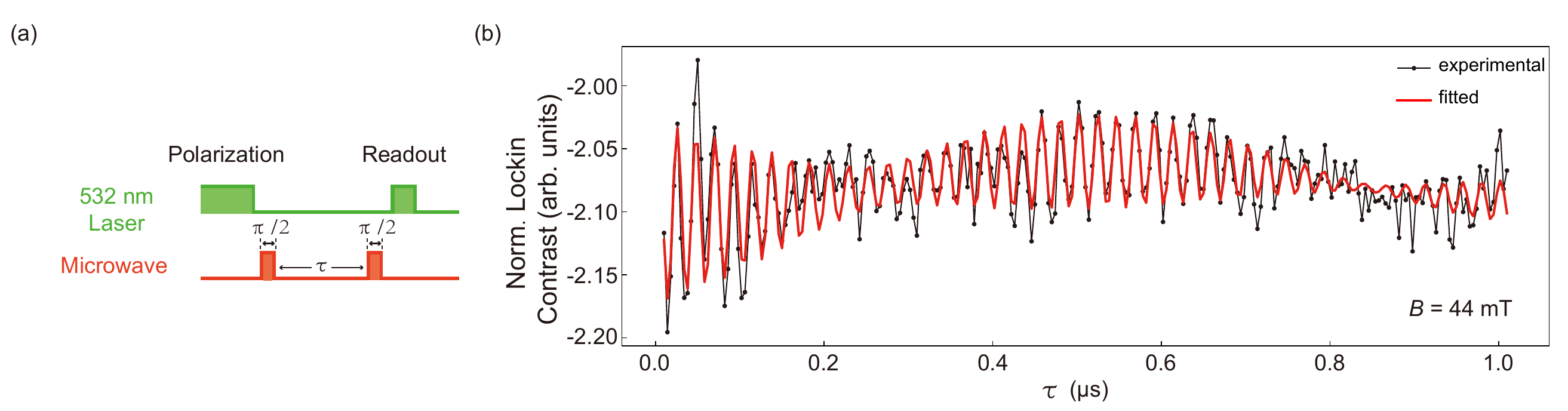}
\caption{Ramsey interference. (a) Ramsey pulse sequence with $\frac{\pi}{2}-\tau-\frac{\pi}{2}$. 
(b) Ramsey result at 44-mT magnetic field driven by 2200-MHz microwave starting from $\tau=10$ ns. The red line is the fitting result \textcolor{blue}{using $\sum_{i=1}^{3} a_ie^{-\tau/T_i}\cos(2\pi f_i\tau+\phi_i)+b$}. Three frequencies are observed and two of them form a clear beat. The distances between the adjacent frequencies are both around $\sim$45 MHz. $T_2^*$ corresponding to these three frequencies are $0.665\pm 0.108$ $\mu$s, $2.500\pm 2.160$ $\mu$s and $1.448\pm 0.841$ $\mu$s, respectively.}
\label{Fig4}
\end{figure*}

\textbf{Rabi oscillations.} The next step is to coherently manipulate the defect spin for which the Rabi oscillation is a key tool. Here, we utilize the two-level electronic states $m_s=\{-1,0\}$ as the electronic-spin qubit to perform the coherent control. The time diagram of the MW and the laser pulse sequences are shown in Fig. 2(a). After a long excitation-laser pulse, the spin is polarized to $m_s=0$ state, then a MW pulse with a length of $\tau$ is applied to rotate the spin, followed by a readout laser pulse. The detected Rabi oscillations are shown in Fig. 2(b). At magnetic field of $B=0$ mT, we see a standard decaying Rabi oscillation, but we also observe a tiny decay of background. By varying the MW power, we derive the linear dependence of the fitted Rabi frequency versus the square root of MW power $\sqrt{P}$ (see Fig. 2(c)), which indicates the validity of our Rabi results. Remarkably, for the results of non-zero magnetic field, we observe the Rabi oscillations containing multiple-frequency components, and the fitted relaxation time of oscillation seems increasing with the magnetic field. 
To reveal the mechanism of the multiple-frequency Rabi oscillation, we carry out the theoretical simulation for the spin dynamics of the closed 4-spin $\rm V_B^-$ system. Our simulation results indicate that the detuning of the MW frequency can drive dynamically the $^{14}$N nuclear spins and lead to the $^{14}$N polarization oscillation as shown in Fig. 2(d). Then the MW-driven $^{14}$N nuclear spin dynamics can modulate the Rabi-oscillation curve, as the simulation results shown in Fig. 2(e), where even a slight detuning of 10 MHz of the MW can induce multiple-frequency oscillation curve.
At strong magnetic field, the $^{14}$N nuclear spins tend to be polarized as we demonstrated by our ODMR results, and the relaxation effect seems weaken according to the fitting results of Rabi oscillation in Fig. 2(b) (see Supplementary Information S4). Hence the $^{14}$N nuclear spins tend to be driven by detuning MW at strong magnetic field and lead to the multiple-frequency Rabi oscillations of $\rm V_B^-$ spin. In addition, according to the theoretical simulation results taking many-body spin bath into account, we conclude that the nuclear spin bath surrounding $\rm V_B^-$ is responsible for the decay of Rabi oscillation, while the nearest neighbor $^{14}$N nuclear spins are responsible for the modulation of multiple-frequency Rabi oscillation and the decay of background (see Supplementary Information S7).

\textbf{$T_1$ measurement and spin echo detections.} With the obtained Rabi frequency, we can define $\frac{\pi}{2}$-pulse and $\pi$-pulse. Utilizing a $\pi$-pulse, we can measure the spin-lattice relaxation time $T_1$. The pulse sequence is shown in Fig. 3(a), and the relaxation result at $B=0$ mT is depicted in Fig. 3(b). By fitting this result, we obtain $T_1=16.377\pm 0.416$ $\mu$s. Then we repeat this sequence for various magnetic fields and obtain the results in Fig. 3(c). We find $T_1$ is approximately independent of the magnetic field. Next, we perform the sequence $\frac{\pi}{2}-\frac{\tau}{2}-\pi-\frac{\tau}{2}-\frac{\pi}{2}$ (spin echo, see Fig. 3(d)) to measure $T_2$. At $B=0$ mT, we observe a monotonic decay of contrast shown in Fig. 3(e), and $T_2$ is fitted to be $82.121\pm 2.462$ ns, which is quite short. 
At $B=36$ mT, we find the decayed-contrast curve is complicatedly modulated (see Fig. 3(f)). We cannot fit it well, and the red line is only a guide for eyes. Here we note that, since $T_2$ is quite short, the impact MW-pulse lengths, especially the $\pi$-pulse length ($\sim$26 to 44 ns in this work), cannot be ignored, therefore, the values of the fitted $T_2$ would have inaccuracy, but the order of magnitude can be determined. For $T_1$, which is much longer than the MW-pulse lengths, this is not an issue.

\textbf{Ramsey interference.} We also perform the Ramsey interference experiment on $\rm V_B^-$ spins. The pulse sequence is presented in Fig. 4(a). We observe no oscillations but the monotonic decay at weak magnetic fields (see Supplementary Information S6), which should be caused by the fast relaxation with the nuclear spin bath, corresponding to a short $T_2^*$.
In contrast, when we apply a magnetic field of $B=44$ mT and set the MW frequency at $f_{\text{MW}}=2200$ MHz, we see a multiple-frequency oscillation, in which a beat is clearly recognized, and it is superposed on another slow oscillation. These three frequencies are fitted as $f_{-1}=-44.171\pm 0.039$ MHz, $f_{0}=0.934\pm 0.131$ MHz, $f_{1}=45.872\pm 0.063$ MHz, respectively, and the distance between the adjacent frequencies are $f_0-f_{-1}=45.105\pm 0.136$ MHz$\approx f_1-f_0=44.938\pm 0.145$ MHz.
Similar to the results of Rabi oscillations, the nearest neighbor $^{14}$N nuclear spins are driven by the MW at strong magnetic field and the observed multiple frequencies in Ramsey interference correspond to spin-rotation frequencies in rotating frame on three hyperfine levels. The frequency $f_{0}$ is the detuning between MW and center hyperfine level, as well as $f_{-1}$ and $f_1$ are the MW detuning from two adjacent nonzero hyperfine levels. In addition, the fitted $T_2^*$ of these three hyperfine levels are $0.665\pm 0.108$ $\mu$s, $2.500\pm 2.160$ $\mu$s and $1.448\pm 0.841$ $\mu$s, respectively. It seems that the magnetic field suppresses the spin relaxation of $\rm V_B^-$ center.

\begin{figure*}[tb]
\centering
\includegraphics[width=0.95\textwidth]{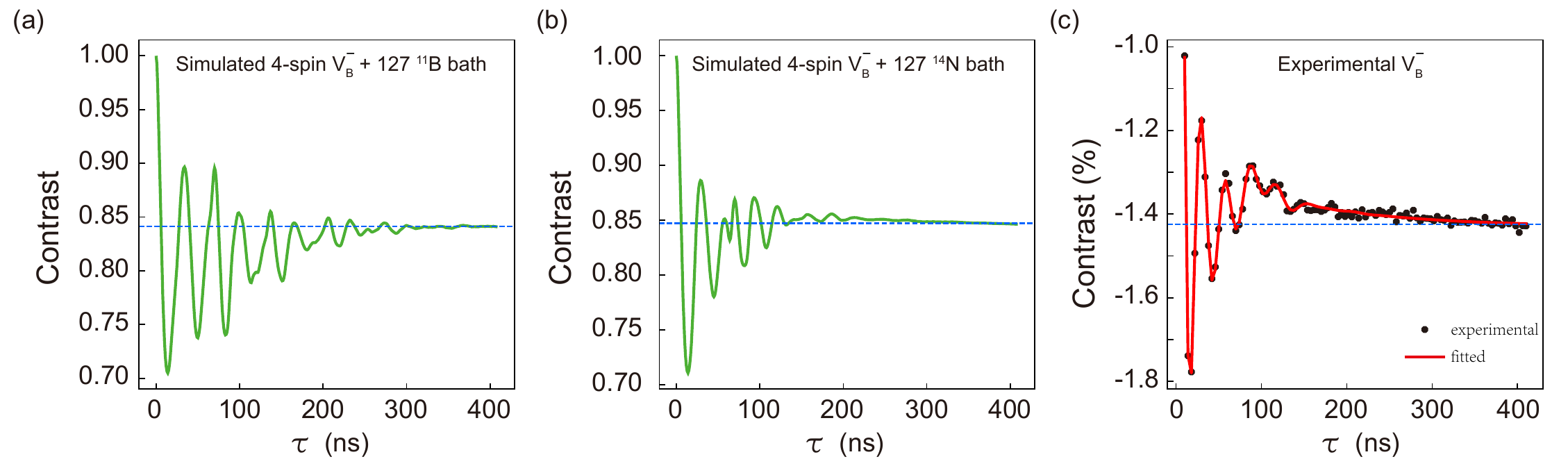}
\caption{Spin dynamics of open $\rm V_B^-$ system with relaxation from the many-body nuclear spin bath at $B=21.7$ mT . (a) and (b) Theoretical Rabi oscillation interacted with the surrounding 127 $^{14}$N or $^{11}$B as the many-body spin bath. The dashed lines are provided as a baseline of long-time background. (c) Experimental Rabi oscillation driven by 2820 MHz microwave at $B=21.7$ mT.}
\label{Fig5}
\end{figure*}

\textbf{Spin-bath relaxation.} To study the spin relaxation mechanism of $\rm V_B^-$ center, we carry out the theoretical simulation for the open $\rm V_B^-$ system, in which many $^{14}$N or $^{11}$B nuclear spins are taken into account to describe the spin bath surrounding the 4-spin $\rm V_B^-$ center in hBN. Cluster approximation and an extended Lindbladian method are used to simulate the $\rm V_B^-$ spin dynamics \cite{ivady2020longitudinal}. Here we consider the spin bath containing 127 $^{14}$N or 127 $^{11}$B nuclear spins and acquire the Rabi oscillation of $\rm V_B^-$ spin with the spin-bath relaxation at a $B=21.7$ mT. The simulation results of the $^{14}$N and $^{11}$B spin bathes are shown in Fig. 5 (a) and (b), respectively. The two simulation results are similar, however, there are also significant differences. In the $^{14}$N spin bath, the Rabi oscillation decays on a longer timescale, while the oscillation's baseline exhibits the obvious positive offset. To verify which spin bath contributes mainly to the relaxation of $\rm V_B^-$ spin, we compare the two theoretical results with the experimental Rabi oscillation at the same magnetic field (Fig. 5(c)). One can see that the major characteristics, like the decay timescale and oscillation's baseline, are captured by the $^{11}$B spin bath. Considering the simulation results of the closed $\rm V_B^-$ center without including a spin bath we observe no decay for long timescale (Fig. 2(e)). Therefore, we conclude that the $^{11}$B-nuclear spin bath is the main source for the relaxation of $\rm V_B^-$ spin dynamics.

\section*{Discussions}

The domination of $^{11}$B nuclear spin for the relaxation effect can be due to the strong hyperfine couplings of $^{11}$B nuclear spins beyond the nearest neighbor $^{14}$N nuclear spins, as well as the possible low polarization of $^{11}$B spin. The inter-nuclear $^{11}$B-$^{11}$B coupling is $\sim$20 times stronger than the $^{14}$N-$^{14}$N coupling, resulting in the fast spin diffusion in the boron sublattice. Besides, we show that the polarization of the nuclear-spin bath in $\rm V_B^-$ can be enhanced by the external magnetic field, then the spin-bath relaxation process like nuclear flip-flop process is suppressed \cite{Yang2014electron,Seo2016quantum,Hanson2008coherent,Shin2013suppression}. Hence at strong magnetic field, the Rabi oscillation and Ramsey interference exhibit enhanced coherent features of the multiple-frequency components and 
are modulated obviously by the MW-driven nuclear spins via the strong electronic-nuclear coupling in $\rm V_B^-$ center. In addition, the asymmetry of ODMR spectra corresponding to the nuclear-spin polarization is also enhanced at the high magnetic fields. It indicates that the nearest neighbor $^{14}$N nuclear spin in $\rm V_B^-$ center could be explored as another controllable spin in hBN.

As the main candidate for spin qubit in vdW material (to date), the coherent operations of defects in hBN based on Rabi oscillation play a crucial role and provide a powerful tool for the design and construction of spin-based vdW-nano-devices, especially when different techniques of vdW heterostructures are combined. Although $T_2$ seems to be still quite short, there will be several methods to improve it.
For example, the short $T_2$ should be related to the relaxation of nuclear spin bath driven mainly by the Fermi contact interaction, and also reduced by the dark electron spin impurities \cite{Haykal2021decoherence}. The higher external magnetic field or decreased neutron-irradiation would be helpful to reduce the coherence relaxation. A suitable annealing on the hBN sample could be also explored since the high-temperature condition can reduce the $\rm V_B^-$ defect number. In addition, putting the sample into a low-temperature cryostat can also reduce the lattice relaxation.

In summary, we demonstrated the room-temperature coherent manipulation and Rabi oscillation of $\rm V_B^-$ spin in hBN, based on which we also detected $T_1$ and performed the spin-echo and Ramsey-interference experiments. We find $T_1$ is almost not affected by magnetic field, however, the results of the Rabi oscillation, spin echo, and Ramsey oscillation are very different under the conditions of weak and relatively strong magnetic fields. To reveal the intrinsic mechanism behind the experimental measurements, we also carried out theoretical simulations on 4-spin $\rm V_B^-$ systems and obtain the theoretical ODMR spectra, nuclear-spin dynamics, and spin-bath relaxation. The theoretical and experimental results are basically consistent with each other, and reveal the strong electronic-nuclear coupling existing in $\rm V_B^-$ center. The $^{14}$N nuclear spins can be driven by the MW and polarized at strong magnetic field. The $^{11}$B nuclear spin bath in hBN should dominate the relaxation of $\rm V_B^-$ spin dynamics. The strong magnetic field can reduce the spin relaxation, and enhance the MW-driven nuclear-spin oscillation to modulate $\rm V_B^-$ spin dynamics.

\emph{Note added}: We found a similar work finished by Gottscholl \emph{et al.} \cite{gottscholl2021room} when we prepare this manuscript. Our results partially coincide with theirs, e.g., $T_1$, which validate the results of each other; but mostly differ from theirs, e.g., the magnetic-field-dependent behaviors, the multiple-frequency oscillations in Rabi and Ramsey results, especially the beat. These different results in both papers complement to each other, and exhibit a more comprehensive knowledge of $\rm V_B^-$ spin defect in hBN.

\section*{Methods}

\textbf{Experimental setup.} A home-built confocal microscopy system combined with a MW system is used for the spin manipulation measurements. A 532-nm laser modulated by an acousto-optic modulator (AOM) is focused on the sample through a 100$\times$ objective (N.A. = 0.9) for the spin initialization and excitation. The photoluminescence with wavelength above 750 nm is collected by the same objective and guided through a multimode fiber to a photoreceiver with high gain ($\sim$10$^{10}$) or an avalanche
photodiode (APD) for signal readout.
The MW generated by a synthesized signal generator is radiated through a 20-$\mu$m diameter copper wire close to the sample or a gold film
microwave stripline plated on the substrate and controlled by a switch. To manipulate the laser and MW, a pulse blaster card is exploited to produce the corresponding electrical pulse sequences. 
A lock-in amplifier or a data acquisition card is finally used to analyze and extract the signal of spin state. For the magnetic field, an electromagnet is placed directly below the sample to generate a $c$-axis field.
See Supplementary Information S1 for more details of the experimental setup.

\textbf{Simulation methodology.} In the theoretical calculations, we study the dynamics of microwave-driven $\rm V_B^-$ spin system by two simulation methods. (1) For the closed 4-spin $\rm V_B^-$ model, we apply the exact time evolution of a 4-spin model, consisting of a center 1-spin electronic spin and its three nearest neighbor 1-spin $^{14}$N nuclear spins. All the hyperfine parameters are taken from our \emph{ab initio} electronic structure calculations. See Supplementary Information S7 for the details of Hamiltonian parameters of $\rm V_B^-$ model. (2) For the open 4-spin $\rm V_B^-$ model, we consider an additional many-body spin bath containing 127 $^{14}$N or $^{11}$B nuclear spins, which are the second and farther neighbor shells of the $\rm V_B^-$ center in hBN. The many-body spin bath interacts with the 4-spin $\rm V_B^-$ system which is described by a cluster approximation combined with an extended Lindbladian method developed in Ref. \cite{ivady2020longitudinal}. In the cluster approximation, the total system is divided into 32 subsystem, and every subsystem contains the 4-spin $\rm V_B^-$ system and one of the nuclear spins in the many-body spin bath. In contrast to method (1), method (2) can induce the relaxation effects in a parameter-free manner. Besides, no additional decoherence and relaxation effects are included beyond the cluster approximation in the simulations. In both methods, the microwave field is added with no approximation, i.e., an oscillating magnetic field with in-plane magnetic field polarization is added to the model to describe the external drive. This allows us to account for the dressing and multi-spin resonances \cite{ivady2020longitudinal}. It should be noted that the theoretical calculations applied in this work are parameter free, i.e., no experimental or adjustable parameters are used in calculations. Hence the theoretical and experimental results are independent of each other.

\section*{Acknowledgments}

This work is supported by the Innovation Program for Quantum Science and Technology (No. 2021ZD0301200), the National Key Research and Development Program of China (No. 2017YFA0304100), the National Natural Science Foundation of China (Grants Nos. 11822408, 11674304, 11774335, 11821404, 11904356, 12174370, and 12174376), the Key Research Program of Frontier Sciences of the Chinese Academy of Sciences (Grant No. QYZDY-SSW-SLH003), the Fok Ying-Tong Education Foundation (No. 171007), the Youth Innovation Promotion Association of Chinese Academy of Sciences (Grants No. 2017492), Science Foundation of the CAS (No. ZDRW-XH-2019-1), Anhui Initiative in Quantum Information Technologies (AHY020100, AHY060300), the Fundamental Research Funds for the Central Universities (Nos. WK2470000026, WK2030000008 and WK2470000028), the Open Research Projects of Zhejiang Lab (No.2021MB0AB02). This work was partially carried out at the USTC Center for Micro and Nanoscale Research and Fabrication. V.I. acknowledges the support from the Knut and Alice Wallenberg Foundation through WBSQD2 project (Grant No. 2018.0071). A.G. acknowledges the Hungarian NKFIH grant No. KKP129866 of the National Excellence Program of Quantum-coherent materials project and the support for the Quantum Information National Laboratory from the Ministry of Innovation and Technology of Hungary, and the EU H2020 project QuanTelCo (Grant No. 862721).

%\acknowledgments
%{\bf  Acknowledgments}

%{\bf Author Contributions}

%{\bf Author Information}

\end{document}